# LanTraP: A code for calculating thermoelectric transport properties with the Landauer formalism

Xufeng Wang[a], Evan Witkoske[a], Jesse Maassen[b], and Mark Lundstrom[a]

[a] Electrical and Computer Engineering, Purdue University, West Lafayette, IN 47907, USA
[b] Department of Physics and Atmospheric Science, Dalhousie University, Halifax, Nova Scotia B3H 4R2, Canada

**Abstract**

A code for calculating the semi-classical thermoelectric and electronic transport coefficients is described. The code uses the Landauer to solve the Boltzmann Transport Equation. In this approach, the central quantity—the transport distribution – is proportional to the distribution of modes times the mean-free-path for backscattering. The distribution of modes (also called channels) can be computed from a given band structure using a so-called band-counting algorithm, which provides fast calculations and offers the potential to rapidly screen DFT band structures. Alternatively, it can be computed from the density-of-states. Good agreement is found when comparing the results obtained using band-counting and established Fourier-based interpolation methods. The code also includes the capability to use energy-dependent scattering times, which is essential for accurate calculation of thermoelectric coefficients. The calculations are benchmarked against the BoltzTraP code for the simple case of constant scattering times.

**Program summary**

*Program title*: LanTraP

*Licensing provisions*: BSD 3-Clause

*Programming language*: MATLAB

*Operating system*: Unix/Linux/PC/Mac

*Distribution format*: ZIP

*Nature of problem*: Calculation of transport properties using Landauer approach

*Solution method*: Fourier interpolation of energy bands coupled with band counting method

## 1. Introduction

Density functional theory (DFT) has become an increasingly powerful tool in thermoelectric material research in recent years [1-6]. It can resolve the material electronic and thermal properties in detail and with accuracy. More and more promising thermoelectric materials are being discovered either from the vast amount of existing DFT material databases [7, 8] or through synthesis by analyzing the material composition with DFT [9-11]. However, calculating various transport properties from DFT is not a trivial task, and much work has been dedicated into solving these numerical issues. For example, band crossings are known to introduce significant numerical issues when calculating transport properties [12, 13]. DFT can under-predict a semiconductors' bandgap, and a scissor-operation is commonly done to correct the bandgap by matching to experimental values [14, 15]. The calculation must be done carefully and k-points near the band edges must be accurately resolved. This is especially important for calculations under non-degenerate conditions or in materials having strong bipolar effects. The degree of "how good" a band structure is also highly depends on details within a narrow window of a few $k_B T$s around the Fermi level. As a result, a band structure giving acceptable accuracy at one temperature may no longer be accurate at another.

The common approach deployed in existing programs is to interpolate the discrete E-k data points obtained from a DFT simulation to a continuous function, so the bands become smooth across the entire Brillouin zone and quantities such as group velocities that depend on derivates of E-k can be readily calculated. In the case of BoltzTraP [16], this continuous function is a Fourier-based, Shankland-Koelling-Wood (SKW) interpolation [17, 18], and in the case of BoltzWann [19], Wannier functions are used instead. However, these interpolations can cost significant computational time, especially when the dataset is large. In addition, if not careful, numerical issues such as ringing, which is an inherent issue in Fourier-based interpolations, may become severe enough to cause errors [18]. This motivated us to seek an alternative method that does not rely on expensive curve-fitting while yielding accurate transport properties. We introduce a software package, Landauer Transport Properties (LanTraP), that uses the Landauer formalism of transport theory [20]. At the center of the Landauer formalism, which is equivalent to the classical Boltzmann theory [21], is a key quantity named the Distribution of Modes (DOM), which enables us to design a fast and efficient algorithm to calculate transport properties without the need of function-based interpolation. In this approach, the kernel of the integrals that determine the transport coefficients, the so-called transport distribution, has a simple, clear physical interpretation; it is proportional to the products of the Distribution of Modes and the mean-free-path for backscattering [22].

The first version of LanTraP was developed by Conrad et al. [23] and was first released in 2013. It is a compact MATLAB code designed to calculate thermoelectric properties given a material's band structure. The functionalities offered by LanTraP are basic, and band structure input obtained from DFT need to be post-processed into a strict format to work with LanTraP. This can create some confusion and a barrier for users to adapt and connect LanTraP with their own programs. It quickly became apparent to us that an improved version of the LanTraP code is needed with new algorithms to increase the accuracy and speed, new functions to handle the data and calculations, and new interfaces with standard DFT software packages. These features will enhance LanTraP and make it suitable to serve the rapidly evolving field of thermoelectric research and efficiently handle the vast amount of research data produced.

## 2. Code implementation

### 2.1. Landauer formalism

At the core of LanTraP is the Landauer formalism used to calculate transport properties. The Landauer formalism has been shown to be equivalent to the classical Boltzmann transport equations [21]. It has the advantage of simplifying some of the transport property calculations with its intuitive approach. As in the case of LanTraP, we take advantage of the Landauer approach and implement a "band-counting" algorithm that can significantly reduce the calculation time of transport properties under some assumptions. We will discuss this in more detail in section 2.2.

The details of the Landauer formalism have been covered by many publications [20, 24-26], and we therefore only state the main results used in LanTraP without derivation. The equations for some key thermoelectric quantities are as follows.

The electrical conductivity is defined as

$$\sigma = \int_{-\infty}^{+\infty} \sigma'(E) dE \tag{1.0}$$

where the differential conductivity is

$$\sigma'(E) = q^2 \Xi(E) \left( -\frac{\partial f_0}{\partial E} \right). \tag{1.1}$$

$\Xi$ is the transport distribution, $f_o$ is the Fermi distribution function, and $q$ is the elementary charge.

The Seebeck coefficient is defined as

$$S = -\frac{1}{qT} \frac{\int_{-\infty}^{+\infty} (E - E_F) \sigma'(E) dE}{\int_{-\infty}^{+\infty} \sigma'(E) dE}. \tag{1.2}$$

The electronic thermal conductivity is defined as

$$\kappa_e = \kappa_0 - T\sigma S^2, \tag{1.3}$$

where the short-circuit electronic thermal conductivity $\kappa_o$ is

$$\kappa_0 = \frac{1}{q^2 T} \int_{-\infty}^{+\infty} (E - E_F)^2 \sigma'(E) dE. \tag{1.4}$$

The key quantity here is the so-called transport distribution function $\Xi$. Its diagonal component is

$$\Xi_{xx}(E) \equiv \sum_{\vec{k}} \upsilon_x(k)^2 \tau_m(E) \delta(E - E_k) \tag{1.5}$$

where an energy-dependent scattering time has been assumed. With the Landauer formalism, the transport distribution has the following form [21]

$$\Xi(E) = \frac{2}{h}(M(E)/A)\lambda(E) \qquad (1.6)$$

where $M(E)/A$ is the number of conduction channels per unit cross-sectional area vs. energy. This quantity is the so-called Distribution of Modes (DOM). The energy-dependent Mean-Free-Path (MFP) for backscattering is also needed; it is defined as [21]

$$\lambda(E) \equiv 2\frac{\langle v_x^2(E)\rangle}{\langle |v_x(E)|\rangle}\tau_m(E) \qquad (1.7)$$

where the quantity, $\langle v_x^2\rangle/\langle |v_x|\rangle$, is an angle-averaged velocity and is computed as a function of energy from the DFT-generated band structure.

To compute the transport coefficients, both $M(E)$ and $\lambda(E)$ are needed. The quantity, $M(E)$, can be obtained directly from the band structure using the band counting technique or indirectly from the Density of States (DOS), and average velocity, $\langle |v_x(E)|\rangle$, both of which can be obtained from the band structure. The Mean-Free-Path (MFP) is also required; it depends on $\langle v_x^2(E)\rangle$ add the energy dependent scattering times, $\tau_m(E)$, which must be specified. One way of specifying $\tau_m(E)$ is built in to the code.).

The DOS accounts for the carrier density distribution with respect to energy, and it can be calculated from a material's full band structure. There are several methods for calculating DOS from DFT in the literature, and each has its own advantages and disadvantages. In LanTraP, we deploy the commonly used tetrahedral integration method for its accuracy and the simple binning method for its speed [27]. Users can choose one of two methods based on the k-grid density of the input band structure.

The DOM accounts for the conduction channels of the carriers, and it can be calculated from a material's full band structure. Two methods are implemented for the DOM calculation. The first method is to compute the DOM from

$$M(E)/A = \frac{h}{4}\langle |v_x(E)|\rangle DOS(E)$$

To provide high accuracy, we use a Fourier-based interpolation and tetrahedron integration method. The second method is aimed at speed and is based on a "band-counting" method derived from the Landauer formalism [22]. This method does not need the velocity at each k point and can thus significantly reduce the calculation time. The algorithms for DOM calculations are presented in section 2.2.

The MFP describes how well these channels can conduct. It is an input to LanTraP and can be obtained through rigorous DFT calculations [28, 29]. One common assumption made in the literature is the constant relaxation time approximation. Under this approximation, the carrier relaxation time is assumed to be constant regardless of the carrier energy. This is a poor assumption if the scattering is dominated by acoustic phonons, which is important for room temperature thermoelectrics [30]. LanTraP incorporates two different scattering options to calculate the MFP. The first option is to have a scattering rate that is proportional to the DOS. For acoustic deformation potential scattering in the elastic limit, the scattering rate is isotropic, equal to the momentum relaxation rate, and proportional to the density of states:

$$\frac{1}{\tau(E)} = \frac{1}{\tau_m(E)} = K_{el-ph} DOS(E),$$ (1.8)

where $K_{el-ph}$ is the electron-phonon coupling parameter. This is a good approximation when acoustic phonon scattering dominates. The second option is to have a constant MFP. This option offers the user a simple and intuitive way to specify scattering. In room temperature TE materials, acoustic deformation potential (ADP) scattering often dominates [31], and for a single parabolic band, it can be shown that the MFP is independent of energy under ADP scattering. The equations for various TE properties are thus greatly simplified. If a TE is nanostructured and has distinguished boundaries for scattering (grain boundaries, voids, defective cracks, etc.), the MFP becomes comparable with the physical size of the distance between these scattering boundaries. Comparing to the constant scattering time approximation, a constant MFP is therefore often a better assumption [32].

## 2.2. Implementation and algorithms

### 2.2.1. LanTraP overall scheme

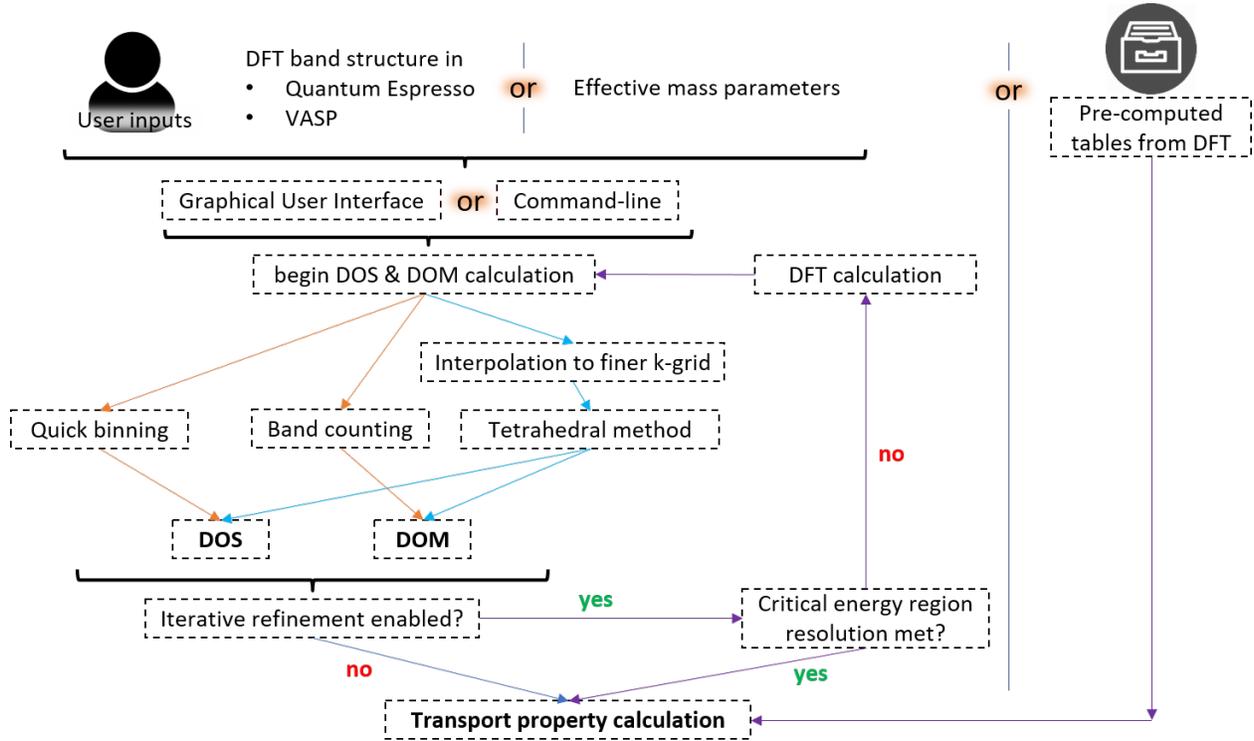

Figure 1. The overall LanTraP program flow chart. Each arrow color represents a calculation route corresponding to an algorithm presented in this section.

The input to LanTraP contains three distinct parts. The first part is the band structure, the second part is the scattering parameters, and the third part is the numerical control parameters.

The band structure input can be analytical or full band. Analytical input allows the user to specify the effective mass tensor, the type (conduction or valence), and the energy minimum (for conduction type) or maximum (for valence type) for each band. This will create a parabolic band structure with multiple bands and is very useful when detailed, DFT-based full band structures are not available. This feature is also useful for comparing results obtained from a complex band structure to that of a corresponding parabolic band.

In case of full band input, LanTraP is able to directly interface with DFT output files from Quantum Espresso and VASP. Information related to crystal structure (lattice constants, symmetries) and E-k are automatically obtained. From here, one can either use the interpolation-based algorithm or the band-counting algorithm to calculate the DOS and DOM. The interpolation-based algorithm implemented in LanTraP is very much like other similar programs including BoltzTraP and BoltzWann. It takes the original E-k and interpolates the energy points onto a finer grid. In LanTraP, we use the Fourier-based

Shankland-Koelling-Wood (SKW) interpolation scheme. For the technical details of this algorithm, please see [16, 18]. Another option is to use the band-counting algorithm, which is discussed in more detail in the next section. It is a non-interpolation-based algorithm that, coupled with the bin counting method, allows fast calculation of DOS and DOM.

The central results of the first part of LanTraP's scheme are the DOS and DOM. Regardless of the input being analytical or full band, the DOS and DOM are passed onto the second part of LanTraP in the same format. Typically, the DOS and DOM calculation from full band, especially with interpolation-based algorithm, is the most time-consuming part. However, the DOS and DOM only depend on the material, so such calculations only need to be completed once before calculating transport properties under various temperatures, doping levels, and scattering parameters. In LanTraP, we provide a small library of DOS and DOM for common TE materials. The user can also save their DOS and DOM calculated by LanTraP into the library for future use. The DOS and DOM are stored in standard MATLAB ".mat" format. Following the equations in section 2.1, various transport properties can be readily obtained from the DOS and DOM.

*2.2.2. Calculating DOM with carrier velocity*

By definition, the DOM along the transport direction (referred to as "x direction" from now on) is related to the carrier group velocity, $v_x$, as follows

$$M(E)/A = \frac{h}{2L} \sum_k |v_x| \delta(E - E_k). \tag{1.8}$$

It is essentially the carrier velocity along the direction of transport multiplied by the DOS. Both the $v_x$ and DOS can be calculated from band structure. As mentioned in section 2.2.1, for calculating DOS, LanTraP can use either tetrahedral integration or simple histogram binning. The calculation of $v_x$ on the other hand can be challenging sometimes. Band group velocity at a certain k-points is related to the derivative of the E-k and requires information of neighboring points. This is especially difficult when the DFT calculated band structure has sparse k-points and band crossings. To get smooth and accurate velocities at all k-points in the Brillouin zone, it is a common practice to fit a continuous function across the entire Brillouin zone for each band, and the velocity can be readily obtained from this fitted function. Existing software such as the popular BoltzTraP [16] and BoltzWann [19] use such interpolation methods to fit the band structure and extract velocities at each k-point. In the case of BoltzTraP, the Shankland-Koelling-Wood (SKW) interpolation method is used, and in the case of BoltzWann, the Wannier function interpolation is used. LanTraP uses the former, and the details of how to obtain $v_x$ using SKW interpolation are provided by BoltzTraP [16] and will not be repeated here.

One issue arises from the minor oscillations in the interpolated curves between data points. Fourier based interpolation techniques have a fundamental issue of ringing due to finite high frequency terms in the Fourier series [17, 33]. Although SKW interpolation improves this issue, minor oscillations may still occur, especially if the interpolation parameters are not optimized. Figure 2 shows such ringing between neighboring points.

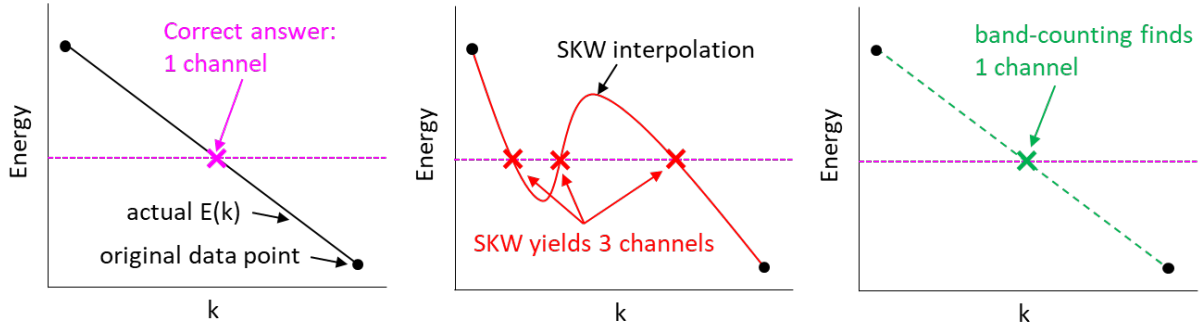

Figure 2. Illustration of the over-estimation of DOM in SKW interpolation method due to ringing.

Ringing impacts the DOS and DOM differently. In the case of DOS calculations, ringing has an effect of broadening and redistributing the states within the energy range of the ringing. The total DOS however does not change. If the ringing occurs within a small energy range, the effect it has on DOS becomes negligible. On the other hand, ringing can have a significant impact on DOM calculations, even when the ringing energy range is small. As illustrated in Figure 2, ringing causes an increase in DOM by the number of oscillations, since each oscillation replaces an original state with two new states—one with positive and the other with negative velocity. This is especially problematic in regions where Gibbs-like multiple ringing occurs after interpolation.

One related issue occurs when treating bands that are "flat". One example can be seen in the conduction band of $Bi_2Te_3$ between high symmetry points Γ and Z [34]. These states have low velocity, and when calculating DOM, resolving these neighboring k-points with comparable energies can cause a similar difficulty as ringing. Overall, one must examine the calculated band structure carefully in order to make sure the DOM calculated is correct and accurate.

*2.2.3. Calculating DOM without carrier velocity*

A significant numerical advantage of the Landauer approach is with the use of the DOM. There are two methods for calculating the DOM: with and without the carrier velocity. We have presented the former—an interpolation-based method in the previous section. The latter so-called "band-counting" method simplifies the calculation of the DOM, and if the MFP is also known, the calculation of transport properties can be done without explicit knowledge of carrier velocity $v_x$.

From the DOM definition shown in Eq. (1.10), one can alternatively show that the DOM can be calculated by simply "counting bands" for a given band structure

$$M(E) = \sum_{k\perp} \Theta(E - E_{k\perp}), \qquad (1.8)$$

where $\Theta$ is a Heaviside function. Eq. (1.11) counts each occurrence of a band having a state with energy E. This so-called "band-counting" method greatly simplifies the numerical calculation of DOM. There is no need to interpolate the band structure and calculate the velocity anymore; one simply needs to count the number of intersections with the energy surface along the transport direction.

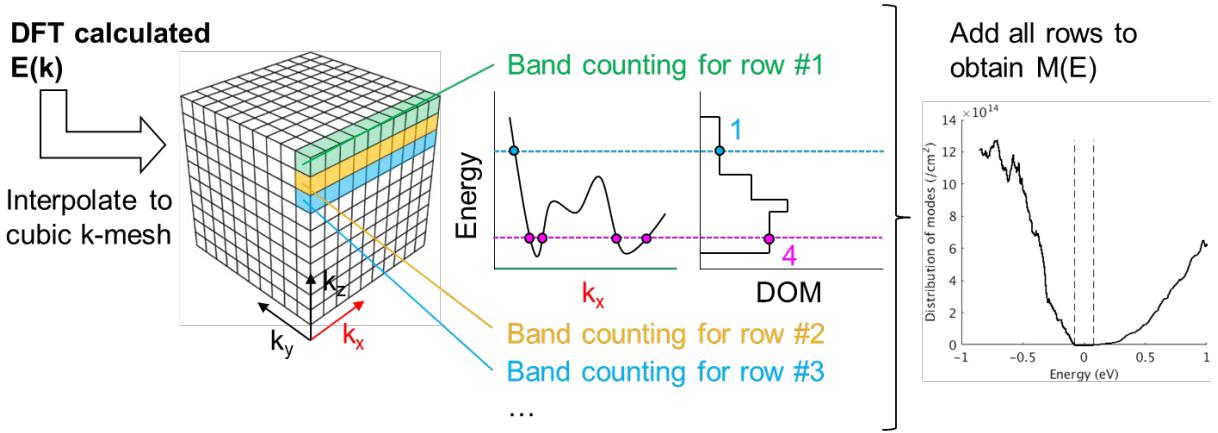

Figure 3. Illustration of the band-counting algorithm calculating the DOM from a DFT calculated E-k.

Figure 3 illustrates how to numerically compute the DOM using the band-counting method from a DFT calculated band structure. The original E(k) is interpolated onto an equivalent k-mesh that aligns with the direction of transport. Notice the interpolation needed here is a simple, nearest neighbor interpolation. It is far less expensive than Fourier or other variants of continuous curve fittings. This linear interpolation can be done in MATLAB simply using, in the case of LanTraP, the *interp3* function. The next step is to count how many times the E(k) along a row of k-points in the direction of transport crosses a given energy E. The result is summed up over all the N×N rows of k-points and divided by the area of the square perpendicular to the transport direction, and this is the DOM at energy E. This counting must be done for all energies to give $M(E)$. An efficient way of doing the counting is to record the energy span of neighboring points along the transport direction and subsequently map these spans into $M(E)$. This way only one round of counting is needed for all the energies and can be formulated into matrix operations, which MATLAB handles effectively.

The main advantage of the band-counting method is the speed gain that comes from omitting the velocity calculations. The curve fitting across the entire Brillouin zone needed to determine velocities can get expensive, especially when the k-point mesh is dense. The lack of velocity information however means one cannot apply the DOS-like scattering rate assumption, since the MFP calculation requires the group velocity at each k-point as shown in Eq. (1.8). Instead, a constant MFP has to be used.

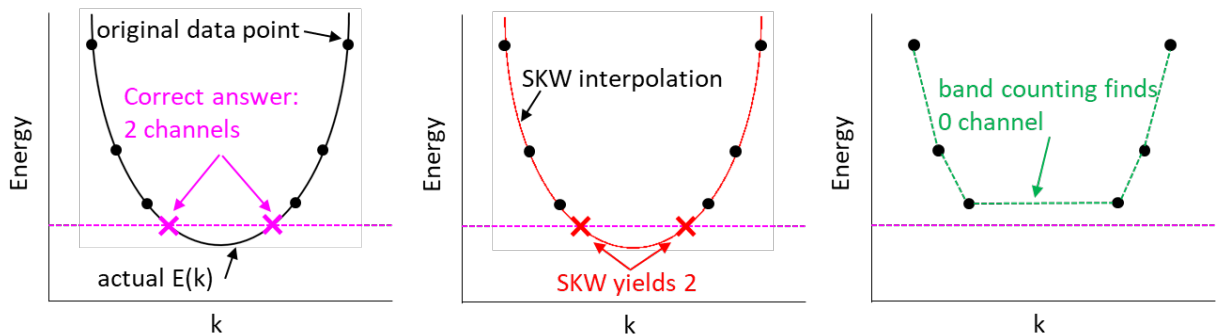

Figure 4. Illustration of the under-estimation of DOM in band-counting method due to inability to smooth out E-k curve.

Numerically, the ringing problem described in the interpolation-based DOM calculation does not exist for the band-counting method, since there is no curve fitting involved. A new numerical issue however arises. As shown in Figure 4, the band-counting method tends to under-predict the DOM. Therefore, the DOM obtained from band-counting represents a lower limit. It is recommended to compare the DOMs obtained from interpolation-based method and band-counting method. If the two DOMs have significant differences, the DFT calculated band structure may not be dense enough, or the numerical interpolation parameters are too strict and cause artificial ringing.

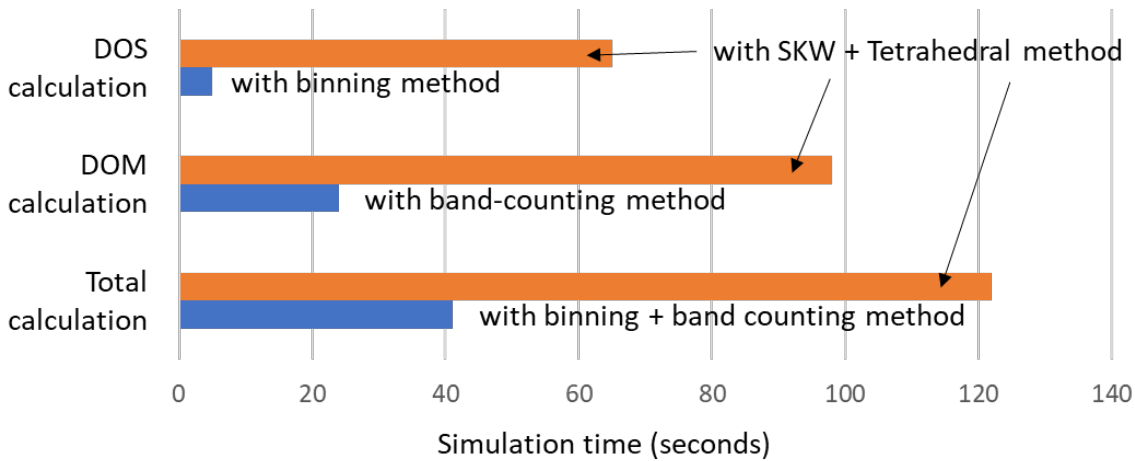

Figure 5. Example speed comparison among the calculation of DOS, DOM, and entire LanTraP simulation with different algorithms. This is done using the silicon example discussed in section 4.1.

The binning method for calculating DOS coupled with the band-counting method for calculating DOM offers a potential for rapid screening of DFT band structures [27]. This combination completely bypasses the need for the time-consuming curve-fitting across the entire Brillouin zone and the calculation of group velocities at each k-point. In Figure 5, we compare the computational time for DOS and DOM calculations in LanTraP with different methods. There is overall, a ~90% time reduction, but it of course comes with a cost of decreased accuracy. In section 4, we present two application test cases and examine the numerical and accuracy issues in more detail.

*2.2.4. Iterative mesh refinement*

The transport properties such as Seebeck coefficient, electrical conductivity, and others are functions of Fermi energy. When evaluating their values using the thermoelectric equations shown in section 2.1, the values of the integrands are non-zero only within several kTs around the Fermi energy due to the $df_0/dE$ term. This critical range of energy is the so-called Fermi window. The Fermi window is only a small portion of the overall band structure energy range which may span several eV. In order to obtain accurate transport property values, a fine k-point mesh is therefore needed. This is especially true for materials with complex band structures. It is always a good idea to double the k-point grid density and check if the calculation results change.

Checking the convergence by doubling the k-point density however can be prohibitively expensive. To address this issue, we propose an iterative scheme to resolve critical energy regions by coupling Quantum Espresso and LanTraP. The scheme is shown in Figure 6. For a certain Fermi level, the Fermi window function is evaluated and a "critical energy range", as one discussed previously, is identified. Within this energy range, energy resolution is checked to see if neighboring energy differences exceed a user-specified value, and if it does, the parallelepipeds containing these k points in the k-grid are recorded. In the end, these parallelepipeds are split into 8 identical sub-parallelepipeds. These newly generated k-points are fed back into Quantum Espresso to obtain new energy solutions. This process iterates until the k-points within the critical energy range gives the desired energy resolution. In Figure 6, we show that this refinement can resolve the correct energy dependence of $E^{1/2}$ for the silicon DOS near the band edge starting from a coarse 40×40×40 Monkhorst-Pack grid.

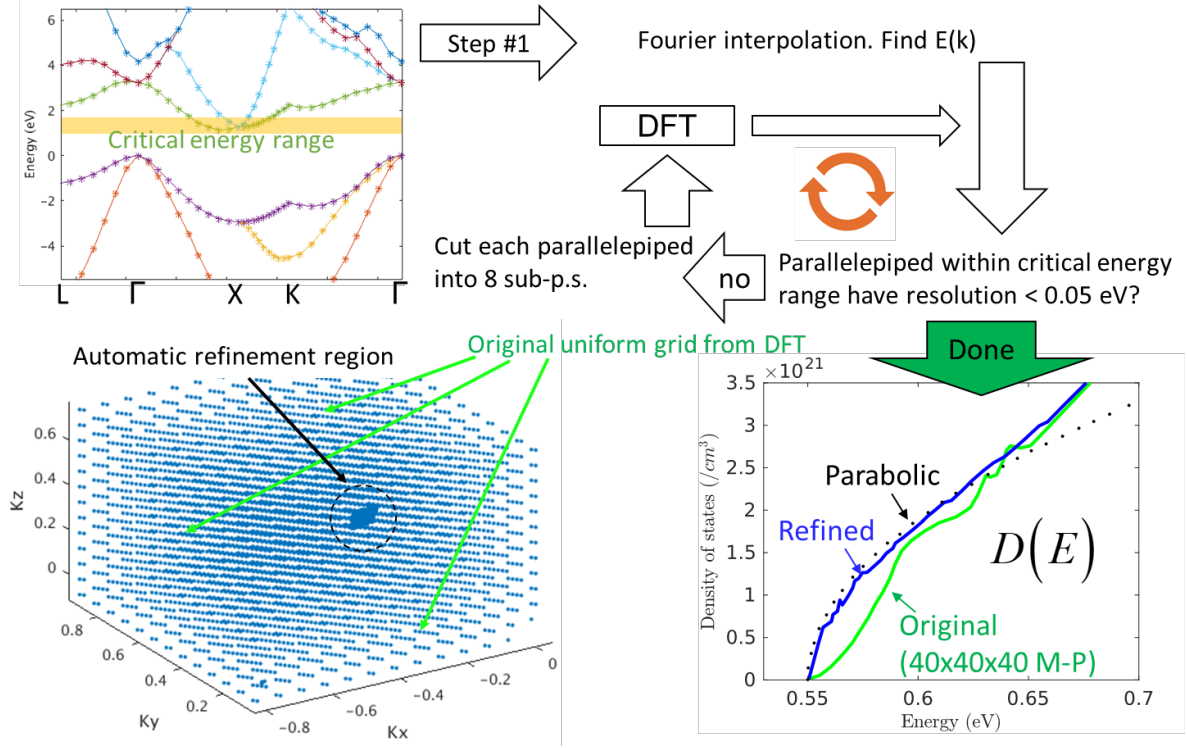

Figure 6. Iterative refinement of E-k within an energy range of interest. An example critical energy range of 0.05 eV is used.

## 3. GUI implementation

A Graphical User Interface (GUI) is available for LanTraP to provide users an easy way to interact with the program. The GUI is built using MATLAB's GUI toolkit [35], thus the GUI window has all standard MATLAB figure features including zoom, export, data probe, and others. The use of a GUI to run LanTraP is optional; one can always use LanTraP as a standard MATLAB function under command-line environment.

The input panel for LanTraP is shown in Figure 7. This panel handles all band structure related inputs, including the import of full band structures from DFT output files, numerical parameters controlling the SKW interpolation of the band structure, and others. The goal is to generate the DOS and DOM to be used in the output panel.

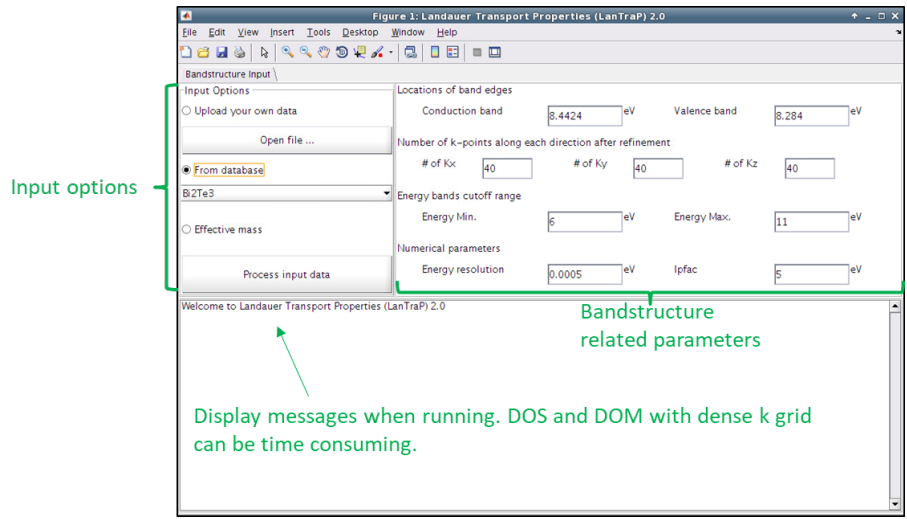

Figure 7. LanTraP GUI input panel. This panel consists of general input options, band structure related parameters, and system display message.

Once "Process input data" is clicked, a new output panel tab is spawned as shown in Figure 8. In this panel, the user can calculate various transport properties under different temperatures, Fermi energy ranges, and scattering models. None of these transport property calculations requires a re-evaluation of the DOS and DOM. Once "Update plots" is clicked, a series of transport property plots are displayed. These plots are natively built into the GUI using standard MATLAB figures, so the user can adjust the figures (zoom, adjust x and y-axis range, etc.) just like regular ones in MATLAB.

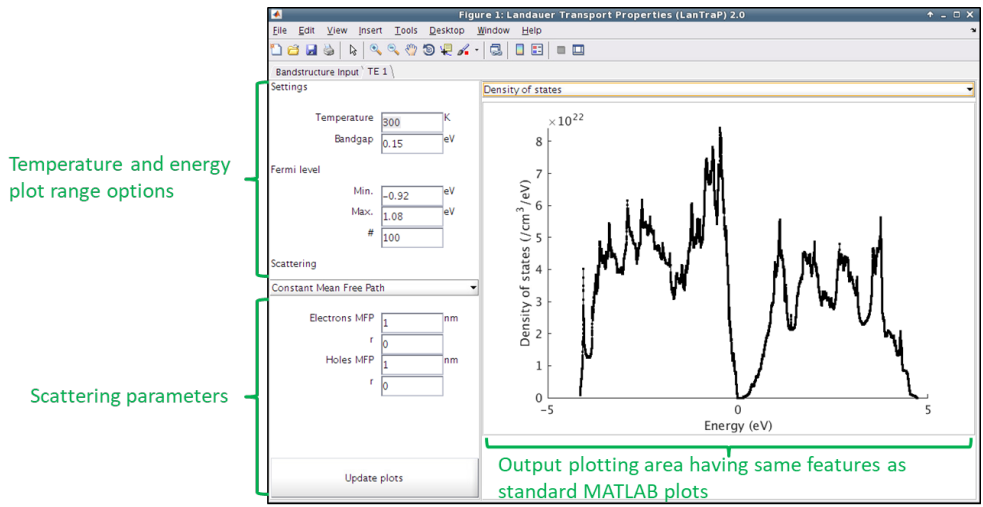

Figure 8. LanTraP GUI output panel. This panel consists of transport property calculation options including temperatures and Fermi energy range, scattering parameters, and plots in MATLAB standard output figure format.

## 4. Application test cases

In this section, we present two examples of LanTraP applications, the analysis of silicon and $Sb_2Te_3$, to illustrate its usage. For both cases, we compare the results obtained with band-counting and interpolation methods to follow up on our discussions on the band-counting method in the previous algorithm section.

### 4.1. Test case: Silicon

In this test case, we use LanTraP to analyze silicon. Silicon is a well-studied material, and its conduction and valence bands near the bandgap are known to be parabolic. The silicon band structure is obtained through first-principles calculations using the Quantum Espresso package [36] with Rappe Rabe Kaxiras Joannopoulos LDA exchange-correlation, norm-conserving pseudo-potential with a 30 Ry plane wave energy cutoff. A 9×9×9 and 100×100×100 Monkhorst-Pack k-mesh is used for self-consistent and non-self-consistent calculations respectively. The obtained band structure is shown in Figure 9(a). The density of states vs. energy, calculated using the tetrahedron method with mesh refinement, is shown in Figure 9(b). Silicon bands are known to be parabolic, we do observe a good parabolic fitting in this case since the DOS closely follows $\sqrt{E}$ near the band edges. The disagreement between parabolic band and full band comes from the higher/lower conduction/valence bands away from the band edges.

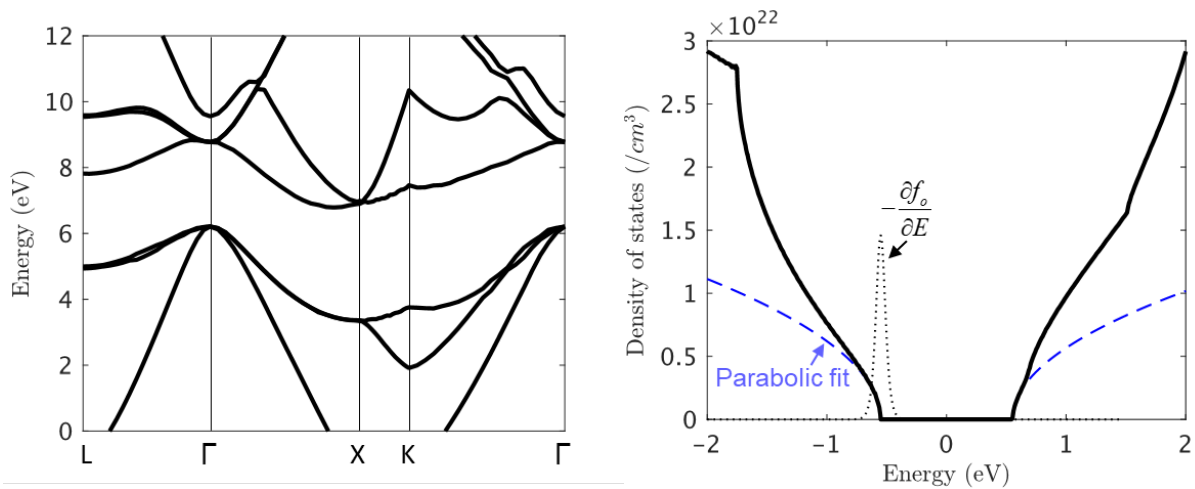

Figure 9. (a) Silicon band structure along all symmetry points obtained from DFT calculation. (b) Density of states of silicon vs. energy (solid black line) near the band gap region with parabolic band fitting applied to both conduction and valence bands. The Fermi window function (dotted black line) is also shown centered around valence band edge at 300 K.

The calculated DOM is shown in Figure 10. As expected, for a parabolic band, the DOM varies linearly with $E$. Here, we compare the DOMs calculated between the SKW method and the band-counting method. The k-point mesh we used is dense, and silicon bands are parabolic with smooth variations. The interpolation into an even denser grid using the SKW method yielded little benefit. This is evident from the comparison with the band-counting method, which uses the original band structure obtained from DFT calculation. The two methods show good agreement overall. The minor differences observed are caused mostly by the numerical inaccuracies in velocity calculations, as discussed in section 2.2.1, and this difference narrows as the k-point mesh becomes denser.

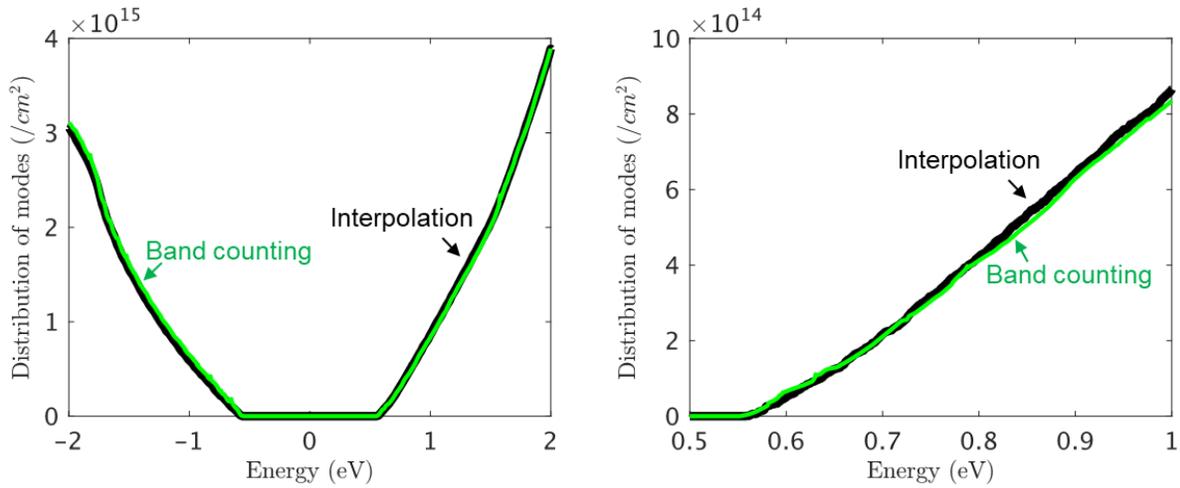

Figure 10. (a) Distribution of modes (DOM) vs. energy for silicon near the band gap region calculated using the Fourier-based interpolation method (thick black line) and the band-counting method (thin green line). (b) DOM vs. energy zoomed in to a narrow energy range (0.5 eV) near the conduction band edge.

Since group velocity information at each k-point is unavailable by using the band-counting method, a constant MFP is used to calculate various transport properties. In Figure 11(a), electrical resistivity vs. electron doping density is plotted using various methods. Under the assumption of a constant MFP, both the SKW and band-counting methods yield similar results. This confirms the small disagreement observed in the DOM in Figure 10 yields only minor differences in the calculated transport properties. This is also observed with the Seebeck coefficient as shown in Figure 11(b). The DOS scattering however makes a noticeable difference when silicon is degenerate. This occurs because the Fermi level is deep into the conduction band and the non-parabolicity becomes significant enough to deviate the MFP from a constant value. LanTraP users thus have to make the judgement on which scattering model to use based on the problem.

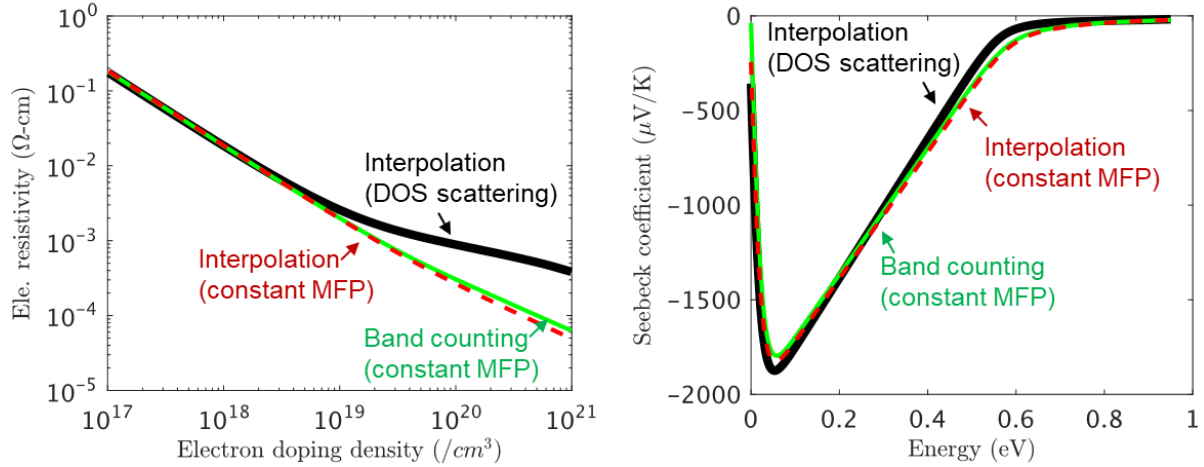

Figure 11. (a) Electrical resistivity vs. electron doping density and (b) Seebeck coefficient vs. energy obtained with the Fourier-based interpolation method under DOS scattering assumption (thick black line), Fourier-based interpolation method under constant mean-free-path assumption (dash red line), and band-counting method under constant mean-free-path assumption (thin green line).

*4.2. Test case: $Sb_2Te_3$*

In this test case, we use LanTraP to analyze a typical bulk thermoelectric material, $Sb_2Te_3$, using a full DFT band structure. Quantum Espresso [36] is used for these $Sb_2Te_3$ DFT calculations. Both Sb and Te used full relativistic PAW pseudo-potentials with a 40 Ry plane wave energy cutoff and spin-orbit coupling. An 8×8×8 and 20×20×20 Monkhorst-Pack k-point mesh is used for self-consistent and non-self-consistent calculations respectively. The calculated band structure is shown in Figure 12(a), and the DOS is shown in Figure 12(b). DFT results are known to under-predict the material bandgap, so we apply a scissor operation and adjust the bandgap of $Sb_2Te_3$ to match the experimental value at 0.2 eV. The bandgap is an input to LanTraP; once specified, LanTraP can locate the conduction and valence band edges and apply the scissor operation to adjust the bandgap automatically. Comparing to silicon, shown in section 4.1, $Sb_2Te_3$ has a more complex band structure and is highly non-parabolic. In addition, the k-point mesh is sparser than what we used for silicon—a challenge due to long computational times that often occurs in rigorous DFT calculations of complex materials.

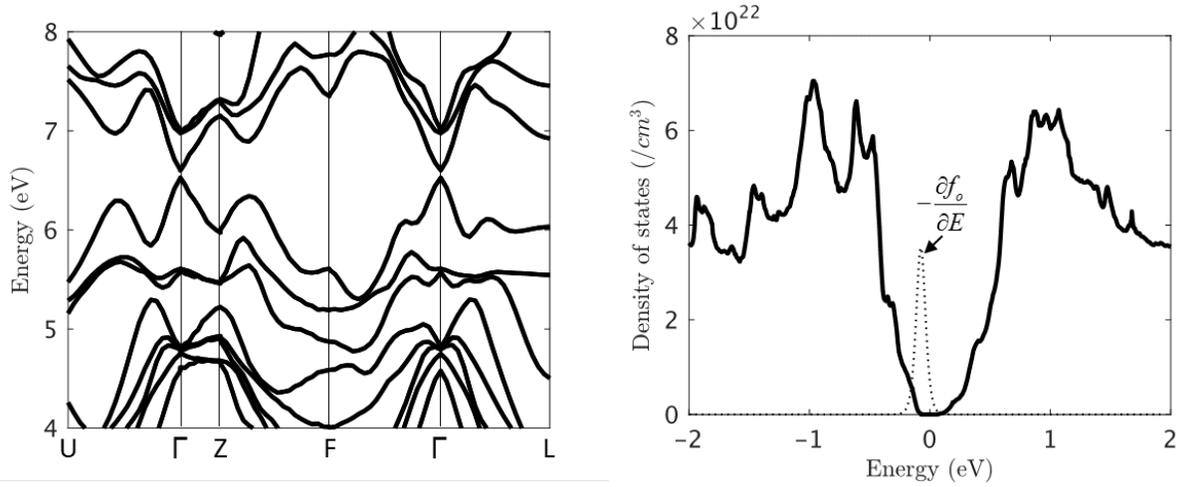

Figure 12. (a) Sb$_2$Te$_3$ band structure along all symmetry points obtained from DFT calculation. (b) Density of states of Sb$_2$Te$_3$ vs. energy (solid black line) near the band gap region with parabolic band fitting applied to both conduction and valence bands. The Fermi window function (dotted black line) is also shown centered around valence band edge at 300 K.

Despite the complex band structure and sparse k-point grid, SKW and band-counting methods yield similar DOM as shown in Figure 13. The difference between the two is more significant than that seen in silicon. For example, Figure 13(b) shows a part of the DOM near the valence band edge, and we can observe that, as discussed earlier in section 2.2.3, the band-counting method tends to under predict the DOM.

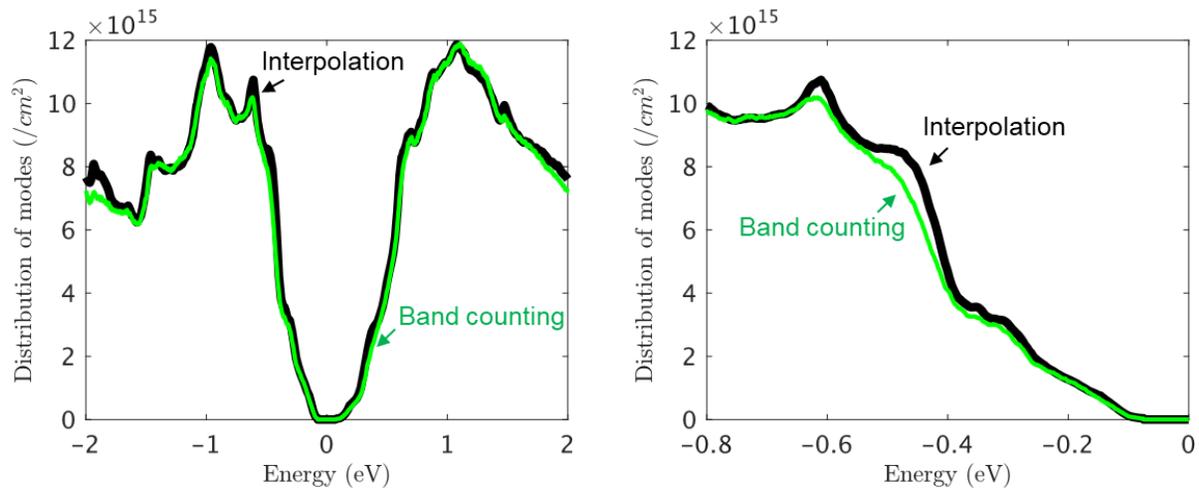

Figure 13. (a) Distribution of modes (DOM) vs. energy for Sb$_2$Te$_3$ near the band gap region calculated using the Fourier-based interpolation method (thick black line) and the band-counting method (thin green line). (b) DOM vs. energy zoomed in a narrow energy range (0.8 eV) near the valence band edge.

Although the band-counting method cannot calculate group velocity at each k-point, LanTraP offers an option to apply a quick SKW-like interpolation to derive the group velocity and then subsequently uses band-counting to calculate DOM. The advantage of this option is to save time by using SKW interpolation without refining/adding addition k-points, while enjoy the band-counting method to rapidly calculate DOM. The SKW interpolation part can still take a significant amount of time. However, this hybrid approach is very useful when a constant MFP is insufficient and the DOS scattering must be used instead.

Figure 14 shows the transport properties calculated with both the SKW interpolation and band-counting methods using a constant MFP and DOS scatterings. First of all, using DOS scattering, both the electrical resistivity and the Seebeck coefficient calculated with the SKW and band-counting methods agree well. As seen in the case of silicon earlier, constant MFP and DOS scattering assumptions produce significantly different results. This is because $Sb_2Te_3$ is highly non-parabolic, so the constant MFP assumption is not suitable. The complexity of some material band structures can lead to significant differences with parabolic band results. This observation led us to use LanTraP as a tool to investigate the Wiedemann-Franz law in complex thermoelectric materials, with the detailed results published elsewhere [37].

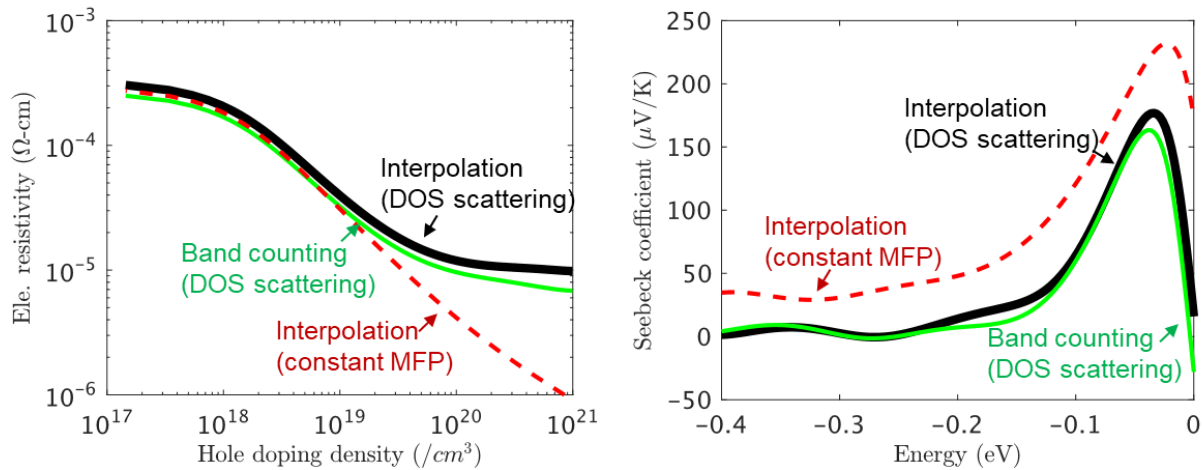

Figure 14. (a) Electrical resistivity vs. electron doping density and (b) Seebeck coefficient vs. energy obtained with the Fourier-based interpolation method under DOS scattering assumption (thick black line), Fourier-based interpolation method under constant mean-free-path assumption (dash red line), and band-counting method under constant mean-free-path assumption (thin green line).

## 5. Conclusions

In this paper, we have presented the implementation and applications of LanTraP—a code for calculating the transport properties of materials using the Landauer formalism. The Landauer equations are equivalent to the Boltzmann equations, and offer an alternative route of calculating transport properties via the distribution of modes. Under the assumption of a constant mean-free-path, we have shown that the calculation of DOM can be greatly simplified through the band-counting algorithm. This opens new opportunities for rapid screening of DFT band structure databases for material discoveries and optimizations.

**Acknowledgement**

This work was partially supported by the Defense Advanced Research Projects Agency (Award No. HR0011-15-2-0037). JM acknowledges support from NSERC (Discovery Grant RGPIN-2016-04881) and Compute Canada.